\documentclass[journal,10pt]{IEEEtran}

\IEEEoverridecommandlockouts %showing the thanks at the left bottom of the first page
\usepackage{epsfig,latexsym}
\usepackage{subfigure}
\usepackage{float}
\usepackage{indentfirst}
\usepackage{amsmath}
\usepackage{amssymb}
\usepackage{cite}
\usepackage{bm}
\usepackage{url}
\usepackage{lastpage}
\usepackage{color}
\usepackage{fancyhdr}
\usepackage{enumerate}
\usepackage{algorithm}
\usepackage{algorithmic}
\usepackage{multirow}
\usepackage{setspace}

\usepackage{epstopdf}

\usepackage[square, comma, sort&compress, numbers]{natbib} %use continus references
\normalsize

\begin{document}
\title{Quantization Design for Resistive Memories with Multiple Reads}
 % <-this % stops a space

\author{Zhen Mei,~\IEEEmembership{Member,~IEEE}, Kui Cai,~\IEEEmembership{Senior Member,~IEEE}, Long Shi,~\IEEEmembership{Senior Member,~IEEE}, and Jun Li,~\IEEEmembership{Senior Member,~IEEE}% <-this % stops a space

%\author{Author 1, Author 2, Author 3, and Author 4% <-this % stops a space
\thanks{
This work was supported by the National Natural Science Foundation of China under Grant 62201258, by the open research fund of National Mobile Communications Research Laboratory, Southeast University (No. 2023D12), by the Jiangsu Specially-Appointed Professor Program 2021, and by Singapore Ministry of Education Academic Research Fund Tier 2 T2EP50221-0036 and RIE2020 Advanced Manufacturing and Engineering (AME) programmatic grant A18A6b0057.  ({\it{Corresponding author: Long Shi}}). 

Zhen Mei, Long Shi and Jun Li are with the School of Electronic and Optical Engineering, Nanjing University of Science and Technology, Nanjing 210094, China (email: meizhen@njust.edu.cn; slong1007@gmail.com; jun.li@njust.edu.cn). Zhen Mei is also with the National Mobile Communications Research Laboratory, Southeast University, Nanjing 210096, China.  

%Zhen Mei is with the School of Electronic and Optical Engineering, Nanjing University of Science and Technology, Nanjing 210094, China. He is also with the National Mobile Communications Research Laboratory, Southeast University, Nanjing 210096, China. (email: meizhen@njust.edu.cn)  
%
%Long Shi and Jun Li are with the School of Electronic and Optical Engineering, Nanjing University of Science and Technology, Nanjing 210094, China (email: slong1007@gmail.com; jun.li@njust.edu.cn).

Kui Cai is with the Science and Math Cluster, Singapore University of Technology and Design, 487372, Singapore (email: cai\_kui@sutd.edu.sg).

%(email: meizhen@njust.edu.cn; slong1007@gmail.com; jun.li@njust.edu.cn)
%Author 1, Author 3, Author 4 with affiliation 1.

%Author 2 with affiliation 2.

} % <-this % %stops a space
}

\maketitle

\begin{abstract}
Due to the crossbar array architecture, the sneakpath problem severely degrades the data integrity in the resistive random access memory (ReRAM).  In this letter, we investigate the channel quantizer design for ReRAM arrays with multiple reads, which is a typical technique to improve the data recovery performance of data storage systems. Starting with a quantized channel model of ReRAM with multiple reads, we first derive a general approach for designing the channel quantizer, for both single-bit and multiple-bit quantizations. We then focus on the single-bit quantization, which is highly suitable for practical applications of ReRAM. In particular, we propose a semi-analytical approach to design the multiple-read single-bit quantizer with less complexity. We also derive the theoretical bit-error probability of the optimal single-bit detector/quantization as the benchmark. Results indicate that the multiple-read operation is effective in improving the error rate performance of ReRAM. Moreover, our proposed multiple-read detector outperforms the prior art detector and achieves the performance of the optimal detector.

\end{abstract}


\section{Introduction}
\IEEEPARstart{E}{emerging} non-volatile memory (NVM) memories such as phase change memory
(PCM), spin-torque transfer magnetic random access memory (STT-MRAM) and resistive random access memory (ReRAM) feature nanoseconds read/write access time, low-energy consumption, long data retention time and almost unlimited endurance \cite{yu2016emerging, mei2018magn}. Compared with PCM and STT-MRAM, ReRAM has greater potential to offer huge density storage due to its simple crossbar structure \cite{strukov2008missing}. However, this simple structure will lead to a poor isolation between different memory cells, which severely affects the reliability of data storage. One fundamental problem that degrades the error rate performance is called the {\it{sneak-path}} problem. In a ReRAM cell, the stored bits of ``0" and ``1" are represented by the high and low resistance states, respectively. During reading, a voltage is applied to the cell and the measured current will determine if the cell is in the high or low resistance state. Sneak paths are paths parallel to the measurement path, and the current going through these paths will distort the reading process of the target cell, leading to an incorrect measurement of the resistance state.

%Moreover, this effect is more significant when reading a high resistance cell, since the low resistance cells in the parallel paths will lower the measured resistance, such that the high resistance state might be incorrectly measured as the low resistance.

%Different approaches have been proposed to address the sneak-path problem \cite{zidan2013memristor, deng2012rram, zidan2014memristor}. In particular, the cell selector has been widely used to eliminate the sneak-path problem \cite{deng2012rram}. The cell selector only allows the current flows in one direction. Hence, by placing a cell selector for each array cell, the sneak paths in the array can be completely mitigated. However, due to the fabrication imperfection and maintenance problem, the cell selector may fail and the sneak-path may occur again. Moreover,

The sneak-path problem has been investigated from information theory and communication theory perspectives \cite{cassuto2016information,  ben2019detection, chen2018coding, song2021performance,nguyen2021two, chen2019variability, zorgui2019non}. Different channel detection and coding schemes were proposed to mitigate the adverse effect of the sneak-path. Specifically, \cite{cassuto2016information, ben2019detection} modeled the sneak-path problem as a type of interference to the cell being being read, based on which \cite{chen2018coding} proposed a diagnal-0 coding schme to reduce the sneak-path interference. \cite{ben2019detection} derived several optimal and sub-optimal detection schemes for the sneak-path interference. In \cite{zorgui2019non}, non-stationary polar codes were proposed to mitigate the sneak-path interference and improve the error rate performance.

%It is a typical technique to improve the data recovery performance of the system.
On the other hand, a system level technique called the read-retry has been widely applied to various data storage systems \cite{cai2015data}.  For ReRAM, a similar scheme named the multiple reads was proposed in \cite{ben2019detection}. However, the low-complexity multiple-read detector proposed by \cite{ben2019detection} is not optimal. Furthermore, it only supports single-bit quantization, and hence the hard-decision decoding (HDD) of error-correcting codes (ECCs). To further improve the error-correction capability, multiple-bit quantization is necessary to facilitate soft-decision decoding (SDD) of ECCs. However, to the best of our knowledge, an information-theoretic design of multiple-bit and multiple-read quantization for ReRAM has not been investigated so far.

In this letter, starting with a quantized channel model of ReRAM with multiple reads, we first derive an information-theoretic approach for designing the quantizer for the ReRAM channel with multiple reads, for both single-bit and multiple-bit quantizations.  We then focus on the single-bit quantization and propose a semi-analytical approach to design the multiple-read single-bit quantizer with less computational complexity. We also derive the bit-error probability of the optimal single-bit detector for the ReRAM channel as the benchmark.

%\begin{table*}[t]
%\centering
%\caption{The sneak-path types of $L=0,1,2,3$.}
%\setlength{\tabcolsep}{1mm}{
%\begin{tabular}{|c|c|c|c|c|c|c|c|c|}
%\hline
%$L$                                      & 0        & 1    & \multicolumn{2}{c|}{2} & \multicolumn{4}{c|}{3} \\ \hline
%$k_r$                                    & 0        & 1    & 1         & 2          & 1    & 2    & 2    & 3 \\ \hline
%$k_c$                                    & 0        & 1    & 2         & 2          & 3    & 2    & 3    & 3 \\ \hline
%$\alpha(\underline{\lambda})$            & $\infty$ & 3    & 2         & 3/2        & 5/3  & 7/5  & 6/5  & 1 \\ \hline
%$\mathcal{A}_{u,v}(\underline{\lambda})$ & 1        & $uv$ &  $\frac{uv(v-1)}{2}$         &   $\frac{uv(u-1)(v-1)}{2}$         &   $\frac{uv(v-1)(v-2)}{6}$    &   $uv(u-1)(v-1)$   &   $\frac{uv(u-1)(v-1)(v-2)}{2}$   &    $\frac{uv(u-1)(v-1)(u-2)(v-2)}{6}$ \\ \hline
%\end{tabular}}
%\end{table*}

\section{Modeling of Quantized ReRAM Channel with Multiple Reads}
\subsection{Preliminary of ReRAM}
\begin{figure}[t]
\centering
\includegraphics[height=1.5in,width=3.2in]{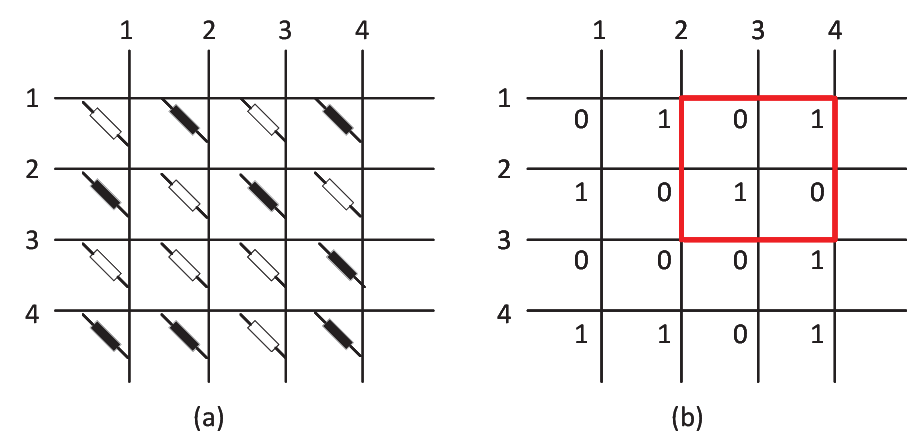}
\caption{(a) A $4\times 4$ crossbar array - white: high resistance cell; black: low resistance cell. (b) The corresponding logical values. }
\label{RRAM_array}
\end{figure}

A resistive crossbar array can be represented by a binary matrix $\mathbf{A}$ with $m$ rows and $n$ columns, with $A_{i,j}$ denoting the bit stored in cell $(i,j)$. Each cell is in one of two resistance states, where the low resistance $R(1)$ represents an input bit ``1" and the high resistance $R(0)$ denotes a bit ``0". During the reading of each cell, the measured resistance will be affected by parallel paths in the array. In particular, when the read cell is in high resistance state, it might be erroneously read as low-resistance if there exists a series of parallel cells having low resistances. In this work, we follow the definition of the sneak-path in \cite{cassuto2016information, ben2019detection}, and consider only sneak-paths of length 3, which is the most dominant case among all sneak-path lengths. Hence, we state that there is a sneak-path to cell $(i,j)$ if $A_{i,c_1} = A_{l_1,c_1} = A_{l_1,j} = 1$, where $c_1\neq j$ and $l_1 \neq i$. This represents the typical case of a sneak-path with three memory cells in parallel to cell $(i,j)$. Fig. \ref{RRAM_array} illustrates an example of $4\times 4$ memory array. As shown in the figure, $(3, 2)\rightarrow (3, 4)\rightarrow (1, 4)\rightarrow (1, 2)\rightarrow (3, 2)$ forms a sneak-path (red line) to the target cell $(3, 2)$. To mitigate the sneak-path interference, the cell selector that allows current to flow only in one direction has been widely adopted. However, the cell selector may fail due to production imperfections. Hence, our model assumes the cell selector failure follows the independent and identically distribution with probability $p_f$ \cite{ben2019detection}.
%Hence, the sneak-path will affect the cell $(i,j)$ only if the cell selector in the cell $(l_1, c_1)$ fails.

The sneak-path interferences are measured by calculating the parallel resistance of the read cell. Due to the crossbar structure of ReRAM, the parallel resistance not only depends on the number of sneak-path paths, but also on the type of sneak-path combination, which is the structure of connections between the resistors involved in the sneak paths. Similar to \cite{ben2019detection}, we define the type of a sneak-path combination as $\underline{\lambda}=(L;k_l,k_c)$, where $k_l$ and $k_c$ denote the numbers of rows and columns that are involved in the $L$ sneak paths. The measured resistance $r$ is decided by the cell's resistance, additive noise and sneak-path interference. Hence, for a given $\underline{\lambda}$, the measured resistance of the cell $(i,j)$ is expressed as
\begin{equation} \label{r}
r = \rho(A_{i,j}, \underline{\lambda}) + \eta,
\end{equation}
where $\rho(A_{i,j}, \underline{\lambda}) = \left( \frac{1}{R(A_{i,j})}+\frac{1}{\alpha(\underline{\lambda})R(1)}  \right)^{-1}$, and $\eta$ is a Gaussian variable that models the combined effect of various noise sources of the memory array, and $\alpha(\underline{\lambda})R(1)$ denotes the parallel resistance of the sneak-path. Different types of $\underline{\lambda}$ and the corresponding values of $\alpha(\underline{\lambda})$ are listed in Table I in \cite{ben2019detection}. In addition, we follow \cite{ben2019detection} and take the values of $R(0)$ and $R(1)$ as $1000\Omega$ and $100\Omega$, respectively. We denote $A_{i,j}$ by $A$ in the following derivations.

According to the channel model given by \eqref{r}, the unconditional transition probability is given by
\begin{equation} \label{trans}
\text{Pr}(r|A=b) = \sum_{\underline{\lambda}'}f_{\eta}\left( r- \rho(b, \underline{\lambda}')\right)p_{\underline{\lambda}}(\underline{\lambda}'),
\end{equation}
where $b\in\{0, 1\}$, $f_{\eta}(\cdot)$ is the Gaussian probability density function (PDF) with zero mean and variance $\sigma_{\eta}^{2}$ and $p_{\underline{\lambda}}(\underline{\lambda}')$ is the probability that a cell has the sneak-path type $\underline{\lambda}'=(L;k_l,k_c)$, given by $p_{\underline{\lambda}}(\underline{\lambda}') = \sum_{u=0}^{m-1}\sum_{v=0}^{n-1}\mathcal{A}_{u,v}(\underline{\lambda}')p_{u,v}p_{L|u,v}$, where $p_{u,v} = \binom{m-1}{u}\binom{n-1}{v}q^{u+v}_{1}(1-q_1)^{m-1-u+n-1-v}$ and $p_{L|u,v} = (p_f q_1)^{L}(1-p_f q_1)^{uv-L} $. $\mathcal{A}_{u,v}(\underline{\lambda})$ is given by Table I in \cite{ben2019detection}.

\subsection{Quantized ReRAM Channel with Multiple Reads}
To design quantizers for ReRAM, a quantized channel model is necessary. To perform a $q$-bit quantization, the channel output $r$ is quantized into $s=2^{q}$ values $\tilde{r}_{0},\tilde{r}_{1},\ldots,\tilde{r}_{s-1} $, where $t_{0}, t_{1} , \ldots, t_{s}$ are the boundaries of quantization intervals with $t_{0}=-\infty$ and $t_{s}=+\infty$, which satisfy $t_{0}<t_{1} < \cdots< t_{s}$. Let $T_{j}=(t_{j}, t_{j+1})$ denote the $j$-th quantization interval, $j=0, 1, \ldots, s-1$. Based on \eqref{trans}, we can obtain the transition probability of the quantized ReRAM channel as
\begin{align}  \label{trans_quan}
\text{Pr}(\tilde{r}_j|A=b) = \text{Pr}(r\in T_{j}|A=b) = \int_{T_{j}}\text{Pr}(r|A=b)dr.
\end{align}
According to \eqref{trans}, for given sneak-path types $\underline{\lambda}'$, $p_{\underline{\lambda}}(\underline{\lambda}')$ are constants and $\text{Pr}(r|A=b)$ becomes a Gaussian mixture PDF. Hence, the cumulative density function (CDF) of $\text{Pr}(r|A=b)$ can be obtained by calculating the weighted sum of the CDF of $f_{\eta}\left( r- \rho(b, \underline{\lambda}')\right)$. Therefore, the transition probability of the quantized ReRAM channel can be rewritten as
\begin{align} \nonumber \label{trans_quan_Q}
&\text{Pr}(\tilde{r}_j|A=b)  = \sum_{\underline{\lambda}'}p_{\underline{\lambda}}(\underline{\lambda}')\int_{T_{j}}f_{\eta}\left( r- \rho(b, \underline{\lambda}')\right)dr\\
&= \sum_{\underline{\lambda}'}p_{\underline{\lambda}}(\underline{\lambda}') \left( Q\left( \frac{t_{j}-\rho(b, \underline{\lambda}')}{\sigma_{\eta}}  \right)-Q\left( \frac{t_{j+1}-\rho(b, \underline{\lambda}')}{\sigma_{\eta}}  \right) \right),
\end{align}
with $Q(x)=\frac{1}{\sqrt{2\pi}}\int_{x}^{\infty}\exp{\left(-\frac{t^{2}}{2}\right)}dt$. To further enhance the performance, we perform $N$ measurements of the same cell's resistance, which is called multiple-read quantization. The readback resistance are given by $\underline{r} = (r(1), r(2), \ldots, r(N))$. The multi-dimensional transition probability is given by \eqref{tran_multi}.

\begin{figure*}
\begin{equation} \label{tran_multi}
\text{Pr}(\tilde{r}_j|A=b)  =  \sum_{\underline{\lambda}'}p_{\underline{\lambda}}(\underline{\lambda}') \int_{T_{j}} \ldots \int_{T_{j}} \frac{\exp \left( -\frac{1}{2\sigma_{\eta}^{2}} \sum_{i=1}^{N} (r(i) - \rho(b, {\underline{\lambda}}'))^2   \right)}{(2\pi\sigma_{\eta}^{2})^{N/2}}  dr(1) \ldots dr(N).
\end{equation}
\hrule
\end{figure*}

\section{General Multiple-Read Quantization Design}
Based on the transition probability given by \eqref{trans_quan_Q} and \eqref{tran_multi}, the single-read and multiple-read quantizations can be designed respectively by maximizing the mutual information (MI). However, for the case with multiple reads, although the transition probability in \eqref{tran_multi} is optimal, the calculation becomes very complex as $N$ increases. Therefore, it is necessary to find a near-optimal but simple method to calculate the transition probability for multiple-read quantization. A natural way is to take the average of $N$ measurements, which reduces the multiple integral in \eqref{tran_multi} to only one integral. Specifically, we denote
\begin{equation} \label{r_bar}
\bar{r} = \frac{1}{N}\sum_{k=1}^{N}r(k),
\end{equation}
where $\bar{r}$ is a random variable. Since $r\sim\mathcal{N}(\rho(A, \underline{\lambda}), \sigma_{\eta}^{2})$, $\bar{r}$ is Gaussian distributed with mean $\mu_{\bar{r}} = \rho(A, \underline{\lambda})$ and variance $\sigma_{\bar{r}}^{2} = \sigma_{\eta}^{2}/N$. Hence, the transition probability is given by
\begin{small}
\begin{align}  \label{trans_quan_Q_multi} \nonumber
&\text{Pr}(\tilde{r}_j|A=b)  \\
&= \sum_{\underline{\lambda}'}p_{\underline{\lambda}}(\underline{\lambda}') \!\left( Q\left( \frac{\sqrt{N}(t_{j}\!-\!\rho(b, \underline{\lambda}'))}{\sigma_{\eta}}  \right)\!-\!Q\left( \frac{\sqrt{N}(t_{j+1}\!-\!\rho(b, \underline{\lambda}'))}{\sigma_{\eta}}  \right) \right).
\end{align}
\end{small}
Note that when $N=1$, \eqref{trans_quan_Q_multi} is reduced to \eqref{trans_quan_Q}, which is the transition probability for single-read quantization. Assuming uniform inputs, the MI of the quantized channel is given by
%\begin{align} \label{I_q1} \nonumber
%I(A;\tilde{r})&= -\sum_{j=0}^{s-1}\text{Pr}(\tilde{r}_{j})\log \text{Pr}(\tilde{r}_{j}) \\
%&+\frac{1}{2}\sum_{j=0}^{s-1}\sum_{A=0}^{1}\text{Pr}(\tilde{r}_{j}|A)\log \text{Pr}(\tilde{r}_{j}|A),
%\end{align}
\begin{equation} \label{I_q1}
\small
I(A;\tilde{r})= -\sum_{j=0}^{s-1}\text{Pr}(\tilde{r}_{j})\log \text{Pr}(\tilde{r}_{j})
+\frac{1}{2}\sum_{j=0}^{s-1}\sum_{A=0}^{1}\text{Pr}(\tilde{r}_{j}|A)\log \text{Pr}(\tilde{r}_{j}|A),
\end{equation}
where $\text{Pr}(\tilde{r}_{j})=\frac{1}{2}\sum_{A=0}^{1}\text{Pr}(\tilde{r}_{j}|A).$ By substituting \eqref{trans_quan_Q_multi} into \eqref{I_q1}, the MI can be calculated for multiple-read quantization. Next, we can optimize the quantization boundaries $t_1, t_2, \ldots, t_{s-1}$ by maximizing $I(A;\tilde{r})$ in \eqref{I_q1}. There are various algorithms that can be employed to find the boundaries, such as the greedy merging \cite{tal2013construct}, dynamic programming (DP) \cite{he2021dynamic} and heuristic algorithms \cite{storn1997differential}. In this work, we employ DP to obtain the optimal quantizer due to its optimality. To perform DP, the ReRAM channel needs to be quantized first. To achieve this, $r$ is uniformly quantized into $h$ intervals $(h\gg s)$ and the threshold is denoted as $u_{i}$ $(0\leq i\leq h)$. Hence, the quantization problem becomes finding $t_1, t_2, \ldots, t_{s-1}$ from $u_{1}, u_{2}, \ldots, u_{h}$, such that the MI is maximized. Note that the MI can be expressed as $I(A;\tilde{r})=H(A)-H(A|\tilde{r})$ and $H(A)$ is a constant. Therefore, maximizing the MI is equivalent to minimizing $H(A|\tilde{r})$, which can be rewritten as $H(A|\tilde{r})=\sum_{j=0}^{h-1}\phi(u_j, u_{j+1})$, where $\phi(u_j, u_{j+1})$ is the partial $H(A|\tilde{r})$ in the range of $[u_j, u_{j+1})$, given by
\begin{equation} \label{MI_psi}
\phi(u_j, u_{j+1})= \frac{1}{2}\sum_{A=0}^{1}\text{Pr}(\tilde{r}_{j}|A)\log \frac{2\text{Pr}(\tilde{r}_{j})}{\text{Pr}(\tilde{r}_{j}|A)}.
\end{equation}
Hence, for $1\leq g\leq o \leq h$, the cost of quantizing $\left\lbrace u_g, u_{g+1}, \ldots, u_0 \right\rbrace $ into one level is given by $\phi(u_g, u_o) =\sum_{j=g}^{o} \phi(u_j, u_{j+1}).$ With this cost function, DP can be used to find the optimal solution $t_1^{*}, t_2^{*}, \ldots, t_{s-1}^{*}$ by recursively solving its subproblems with complexity $O((h-s)^2 s)$. The details of the DP algorithm can be found in Algorithm 1 in \cite{he2021dynamic}.

%First, the ReRAM channel needs to be uniformly quantized into $h$ intervals ($h=1000$ and $h\gg s$) and the quantization boundary is denoted as $u_{i}$ $(0\leq i\leq h)$. Then, the quantization problem becomes finding $t_1, t_2, \ldots, t_{s-1}$ from $u_{1}, u_{2}, \ldots, u_{h}$, such that the MI is maximized. DP can be used to find the optimal values $t_1^{*}, t_2^{*}, \ldots, t_{s-1}^{*}$  with complexity $O((h-s)^2 s)$. The details of the corresponding DP algorithm can be found in Algorithm 1 in \cite{he2021dynamic}.
% Note that the above DP process can be carried out off-line. For practical implementation, the designed quantization boundaries can be used directly by the analog-to-digital converter (ADC).

\begin{figure}[t]
\centering
\includegraphics[height=2.2in,width=3.3in]{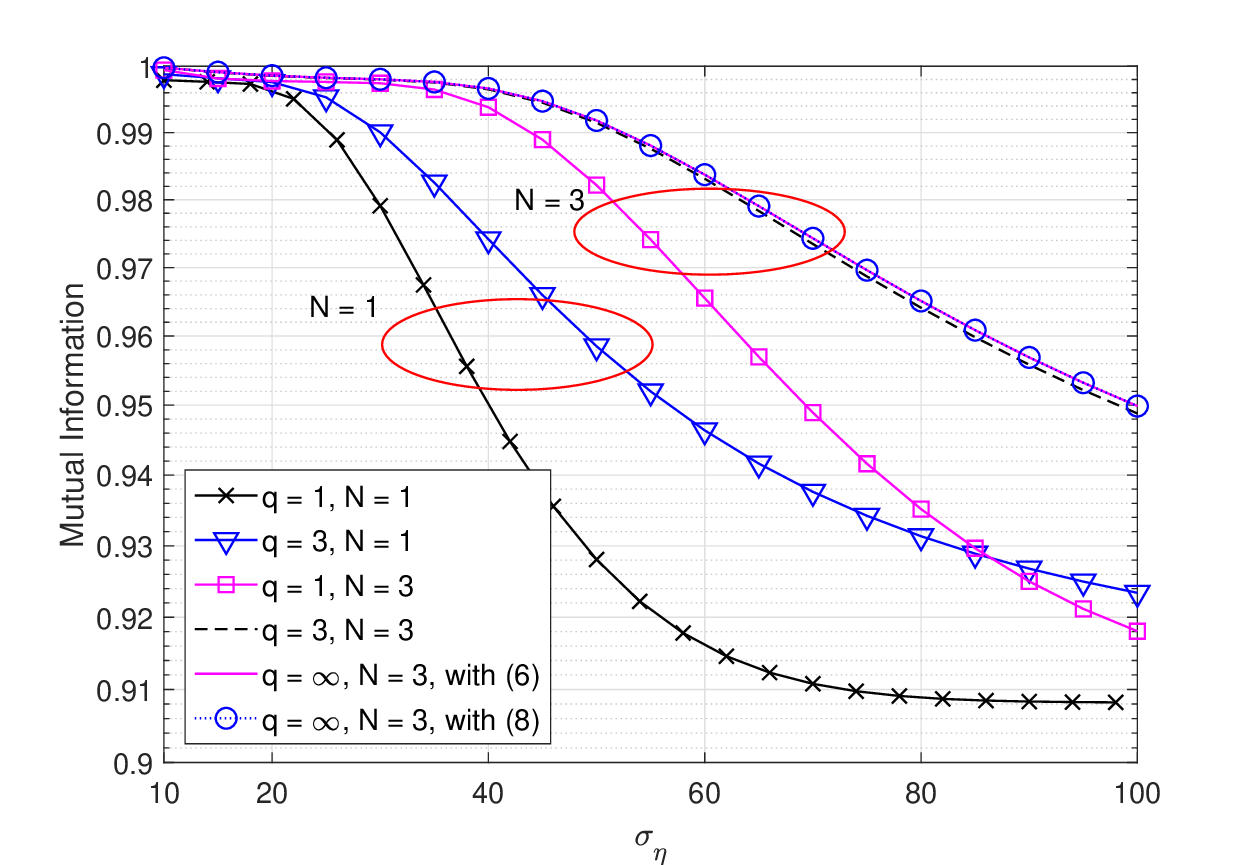}
\caption{The maximum MI of the ReRAM channel with different number of quantization bits ($q=1,3$) and different number of reads ($N = 1, 3$).}
\label{MI_figure}
\end{figure}
Fig. \ref{MI_figure} illustrates the maximum MI of the ReRAM channel with the designed quanitizers, with different number of quantization bits and different number of reads, for an array of size $m = n = 16$ and $p_f=0.001$. Observe that the optimal MI calculated based on \eqref{trans_quan_Q_multi} approaches that calculated according to \eqref{tran_multi}, for the case of $q=\infty$ and $N=3$.  This demonstrates that $\eqref{r_bar}$ is a near-optimal metric to calculate the MI of the ReRAM channel with multiple reads. Furthermore, for the case with single-bit quantization, we have another important observation. That is, by increasing the number of reads, the MI of single-bit quantization can be greatly improved. It can be seen that with $N=3$, single-bit quantization performs even better than 3-bit quantization with $N=1$ for $\sigma_\eta<85$. Hence we can conclude that through introducing more number of read operations, we can significantly improve the error rate performance with single-bit quantization. In the next subsection, we focus on the multiple-read single-bit quantization and propose a design approach that is much faster than the DP.

%Also, with $N=8$, the single-bit quantization achieves larger MI than the 3-bit quantization with $N=3$. Note that the multiple-read detector in \cite{ben2019detection} adopts $N=10$ reads, with which our proposed single-bit quantization can achieve even higher MI.

%given by $\frac{\text{Pr}(r|A=1)}{\text{Pr}(r|A=0)}\overset{\hat{A}=0}{\underset{\hat{A}=1}{\lessgtr}} 1$, where $\hat{A}$ is the detected bit. However,

\section{Multiple-Read Single-Bit Quantization Design}
\subsection{Semi-Analytical Approach to Design the Multiple-Read
Single-Bit Quantizer}
Single-bit quantization is highly suitable for practical applications of ReRAM. It is equivalent to designing a threshold, thus it becomes a threshold detector. The optimum single-bit detector is the MAP detector, which is too complex to be implemented. To reduce the complexity of the MAP detector, sub-optimal detectors have designed in \cite{ben2019detection}. In this section, we propose to design the detectors by maximizing the MI.

%\begin{figure}[t]
%\centering
%\includegraphics[height=1.3in,width=2.3in]{BAC.eps}
%\caption{The equivalent BAC of one-bit quantized RRAM channel.}
%\label{BAC}
%\end{figure}

Given a threshold $t_{1}$, the quantized ReRAM channel only has two possible outputs $\tilde{r}_0$ and $\tilde{r}_1$. After making hard decisions, it is equivalent to a binary asymmetric channel (BAC) whose input is $A$ and output is $\hat{A}$ with $\hat{A}\in \{0, 1\}$. We denote the crossover probabilities of the BAC by $\text{Pr}(\hat{A}=1|A=0)=p_0$ and $\text{Pr}(\hat{A}=0|A=1)=p_1$, respectively. Thus, $\text{Pr}(\hat{A}=0|A=0)=1-p_0$ and $\text{Pr}(\hat{A}=1|A=1)=1-p_1$.

\newcounter{TempEqCnt}
\setcounter{TempEqCnt}{\value{equation}}
\setcounter{equation}{15}
\begin{figure*}[ht]
\begin{equation} \label{map_bep}
P(e|A=b, \underline{\lambda}') = \int_{r_{1}} \int_{r_{2}}\ldots \int_{r_{N}} \frac{\exp \left( -\frac{1}{2\sigma_{\eta}^{2}} \sum_{m=1}^{N} (r_m - \rho(b, \underline{\lambda}'))^2   \right)}{(2\pi\sigma_{\eta}^{2})^{N/2}}  dr_1 dr_2 \ldots dr_N.
\end{equation}
\hrule
\end{figure*}

\setcounter{equation}{\value{TempEqCnt}}
%\begin{equation} \label{th_multiple}
%\bar{r}=\frac{1}{N}\sum_{k=1}^{N}r(k) \overset{\hat{A}=0}{\underset{\hat{A}=1}{\lessgtr}} \tau,
%\end{equation}

For the multiple-read case, threshold detector is given by $\bar{r}=\frac{1}{N}\sum_{k=1}^{N}r(k) \overset{\hat{A}=0}{\underset{\hat{A}=1}{\lessgtr}} \tau$, where $\tau$ is the threshold. The transition probabilities for the multiple-read threshold detector are given by
\begin{small}
\begin{equation} \label{w0_multi}
\text{Pr}(\hat{A}=1|A=0) = \sum_{\underline{\lambda}'}p_{\underline{\lambda}}({\underline{\lambda}}') Q\left( \frac{\sqrt{N}(t_1-\rho(0, \underline{\lambda}'))}{\sigma_{\eta}} \right)
\end{equation}
and
\begin{equation} \label{w1_multi}
\text{Pr}(\hat{A}=0|A=1) =1- \sum_{\underline{\lambda}'}p_{\underline{\lambda}}({\underline{\lambda}}')Q\left( \frac{\sqrt{N}(t_1-\rho(1, \underline{\lambda}'))}{\sigma_{\eta}} \right).
\end{equation}
\end{small}

By substituting \eqref{w0_multi} and \eqref{w1_multi} into \eqref{I_q1}, we can calculate the MI of the single-bit quantized ReRAM channel. Let $t_{1}^{*}$ denote the optimum threshold that maximizes the MI. In this subsection, $t_{1}^{*}$ is calculated by a derivative-based method.

%Since the MI is locally quasi-concave in the range of our interest, we can use bisection search to find $t_{1}^{*}$ instead of using the DP approach, with much lower computational complexity.

To determine $t_{1}^{*}$, we first compute the derivative of \eqref{I_q1} with respect to $t_1$. As the single-bit quantized channel is reduced to the BAC, to simplify the calculation of the derivative, we can employ another expression of the MI, given by
\begin{equation} \label{sym_c}
I(A;\hat{A})=\frac{1}{2}\sum_{\hat{A}=0}^{1}\sum_{A=0}^{1}\text{Pr}(\hat{A}|A)\log\frac{\text{Pr}(\hat{A}|A)}{\text{Pr}(\hat{A})},
\end{equation}
where $\text{Pr}(\hat{A}) = \frac{1}{2}(\text{Pr}(\hat{A}|A=0)+\text{Pr}(\hat{A}|A=1)).$
By substituting \eqref{w0_multi} and \eqref{w1_multi} into \eqref{sym_c}, we can obtain the MI for the BAC. Taking the derivative of \eqref{sym_c}, we can obtain
\begin{align} \label{deriv} \nonumber
\frac{dI(A;\hat{A})}{dt_{1}} &= \frac{p'_{0}}{2}\log\frac{p_{0}(1-p_{0}+p_{1})}{(1-p_{0})(1+p_{0}-p_{1})} \\
&\quad   +\frac{p'_{1}}{2}\log\frac{p_{1}(1+p_{0}-p_{1})}{(1-p_{1})(1-p_{0}+p_{1})} ,
\end{align}
where $p'_{i} (i=0,1)$ is given by $p'_{i} =\frac{(2i-1)\sum_{\underline{\lambda}}p_{\underline{\lambda}}({\lambda}')  }{\sqrt{2\pi}\sigma_{\eta}} \exp\left( -\frac{(t_{1}-\rho(i, \underline{\lambda}))^2}{2\sigma_{\eta}^{2}} \right)$. According to \eqref{deriv}, the first-order derivative of the MI has only one zero point in the range of $[R(1), R(0)]$. Therefore, the MI only has one extrema within this range. Furthermore, it can be verified that the second-order derivative of the MI is smaller than 0 between $200\Omega$ and $1000\Omega$. Therefore, the MI is a locally concave function in the range of our interest and it reaches the maximum at $\frac{dI(A;\hat{A})}{dt_{1}}=0$. Hence, $t_{1}^{*}$ can be computed by using bisection search method.

The complexity of bisection search for $t_{1}^{*}$ is $O(\log N_s)$, where $N_s$ is the number of samples in the range of $[R(1), R(0)]$, and $N_s=128$ has been found to be sufficient. Note that this method is independent of the number of quantization levels. As described in Section III, the complexity of DP-based quantization design is $O((h-s)^2 s)$, where $h$ and $s$ are the initial and target numbers of quantization levels, respectively. In our design, $h$ is set to 1000 to preserve most of the MI. Therefore, it can be seen that the complexity of the proposed semi-analytical approach is much less than the DP for single-bit quantization.

%\begin{align}  \nonumber
%p'_{i} &=\frac{(2i-1)\sum_{\underline{\lambda}}p_{\underline{\lambda}}({\lambda}')  }{\sqrt{2\pi}\sigma_{\eta}} \exp\left( -\frac{(t_{1}-\rho(i, \underline{\lambda}))^2}{2\sigma_{\eta}^{2}} \right).
%\end{align}

\subsection{BEP of the Optimum Multiple-Read Single-Bit Detector}
As a benchmark, we derive an analytical BEP for the multiple-read MAP detectors. The general expression of the BEP is given by
\begin{eqnarray}\label{pee}
P_{b}^{\text{MAP}}=P(A=0)P(e|A=0) + P(A=1)P(e|A=1),
\end{eqnarray}
%\begin{align}  \label{pee}
%P_{b}^{\text{MAP}} &= P(A=0)P(e|A=0) + P(A=1)P(e|A=1) ,
%\end{align}
where $P(A=0)=P(A=1)=0.5$. Here, $e$ is the event that an error occurs and
\begin{equation} \label{peb2}
P(e|A=b) = \sum_{\underline{\lambda}'}P(e|A=b, \underline{\lambda}') p_{\underline{\lambda}}({\underline{\lambda}}'),
\end{equation}
with $b\in\{0, 1\}$ and $\bar{b}=1-b$. For the multiple-read single-bit detector with $N$ measurements, $P(e|A=b, \underline{\lambda}')$ is given by \eqref{map_bep}. The integral limits $\underline{r} = \{r_1, r_2, \ldots, r_N\}$ in \eqref{map_bep} is determined by finding $\underline{r}$ such that
$\text{Pr}(\underline{r}|\bar{b}) \geq \text{Pr}(\underline{r}|b)$, where
\begin{equation} \nonumber
\text{Pr}(\underline{r}|b)=\sum_{\underline{\lambda}'}\frac{p_{\underline{\lambda}}({\lambda}')}{(2\pi \sigma_{\eta}^2)^{N/2}}\exp\left( -\frac{1}{2\sigma_{\eta}^2}\sum_{k=1}^{N}(r(k)-\rho(b, \underline{\lambda}'))^2  \right).
\end{equation}
By substituting \eqref{peb2} and \eqref{map_bep} into \eqref{pee}, we can obtain the final expression of the BEP.

%For single-read detector, $P(e|A=b, \underline{\lambda})$ is given by
%\begin{equation} \label{peb_single}
%P(e|A=b, \underline{\lambda}) = \int_{r:\text{Pr}(r|\bar{b}) \geq \text{Pr}(r|b)}\text{Pr}(r|A=b)dr
%\end{equation}
%By substituting \eqref{peb} and \eqref{peb_single} into \eqref{pe}, we can calculate the theoretical BEP for the single-read MAP detector.

\begin{figure}[b]
\centering
\includegraphics[height=2.2in,width=3.3in]{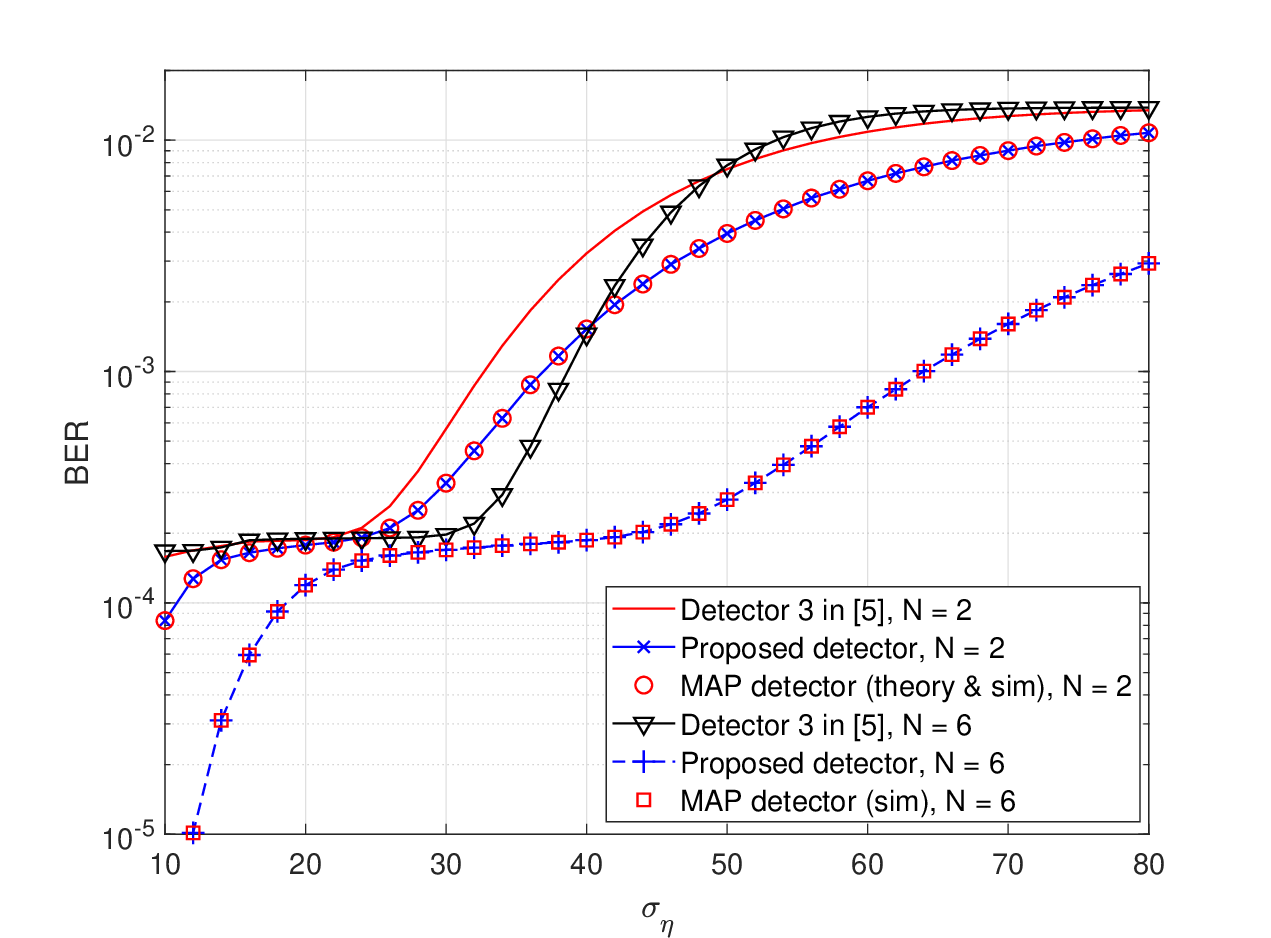}
\caption{The BER comparison of Detector 3 in \cite{ben2019detection} and our proposed detector with multiple-read ($N = 2, 4$) single-bit quantization.}
\label{Uncoded_BER}
\end{figure}

\section{Simulation Results}

\begin{figure}[t]
\centering
\includegraphics[height=2.2in,width=3.4in]{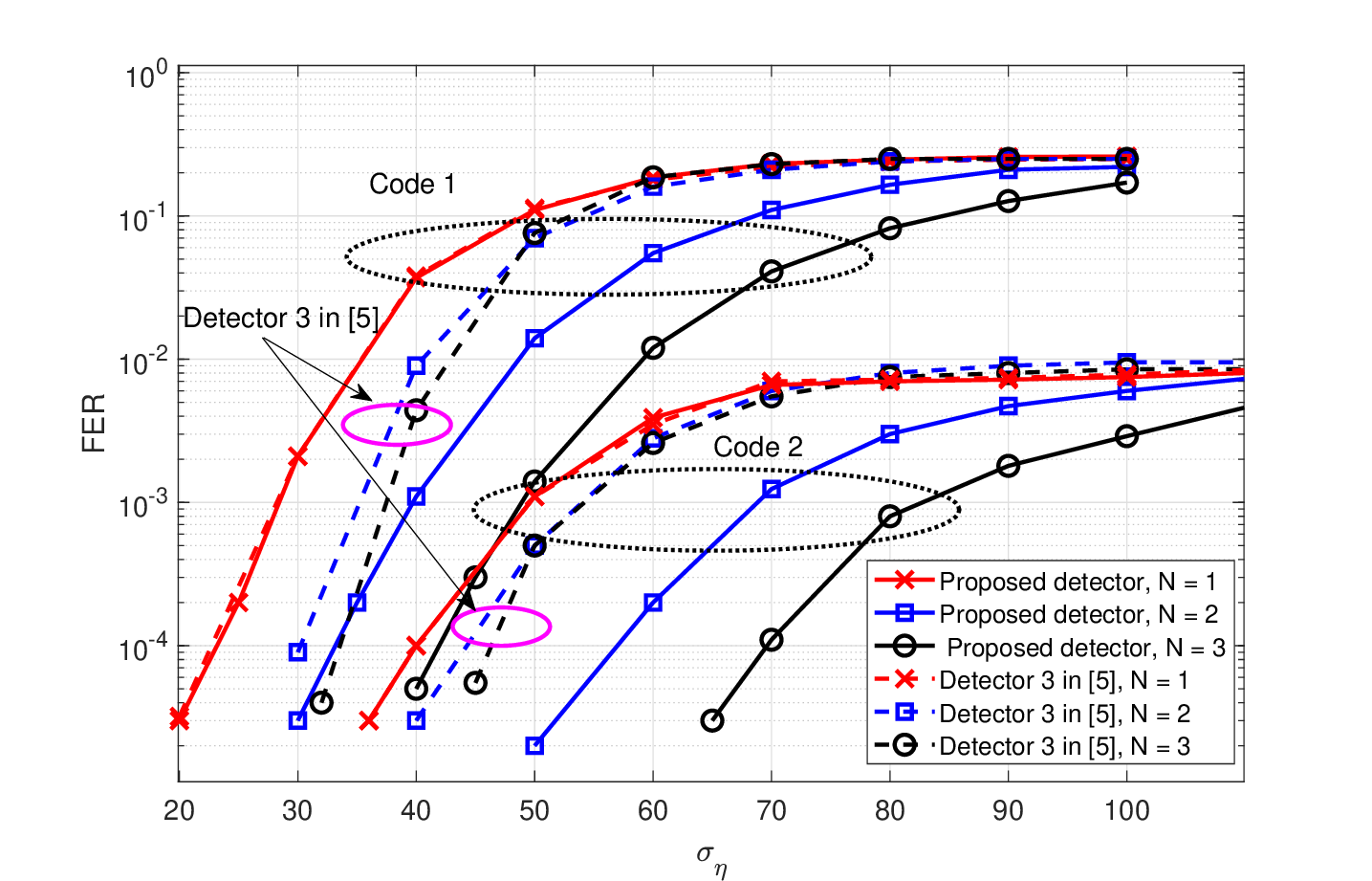}
\caption{The FER of (127, 113) BCH code (Code 1) and (127, 92) BCH code (Code 2) with Detector 3 in \cite{ben2019detection} and our proposed detector with different number of reads and single-bit quantization.}
\label{BCH_FER}
\end{figure}

\begin{figure}[t]
\centering
\includegraphics[height=2.2in,width=3.4in]{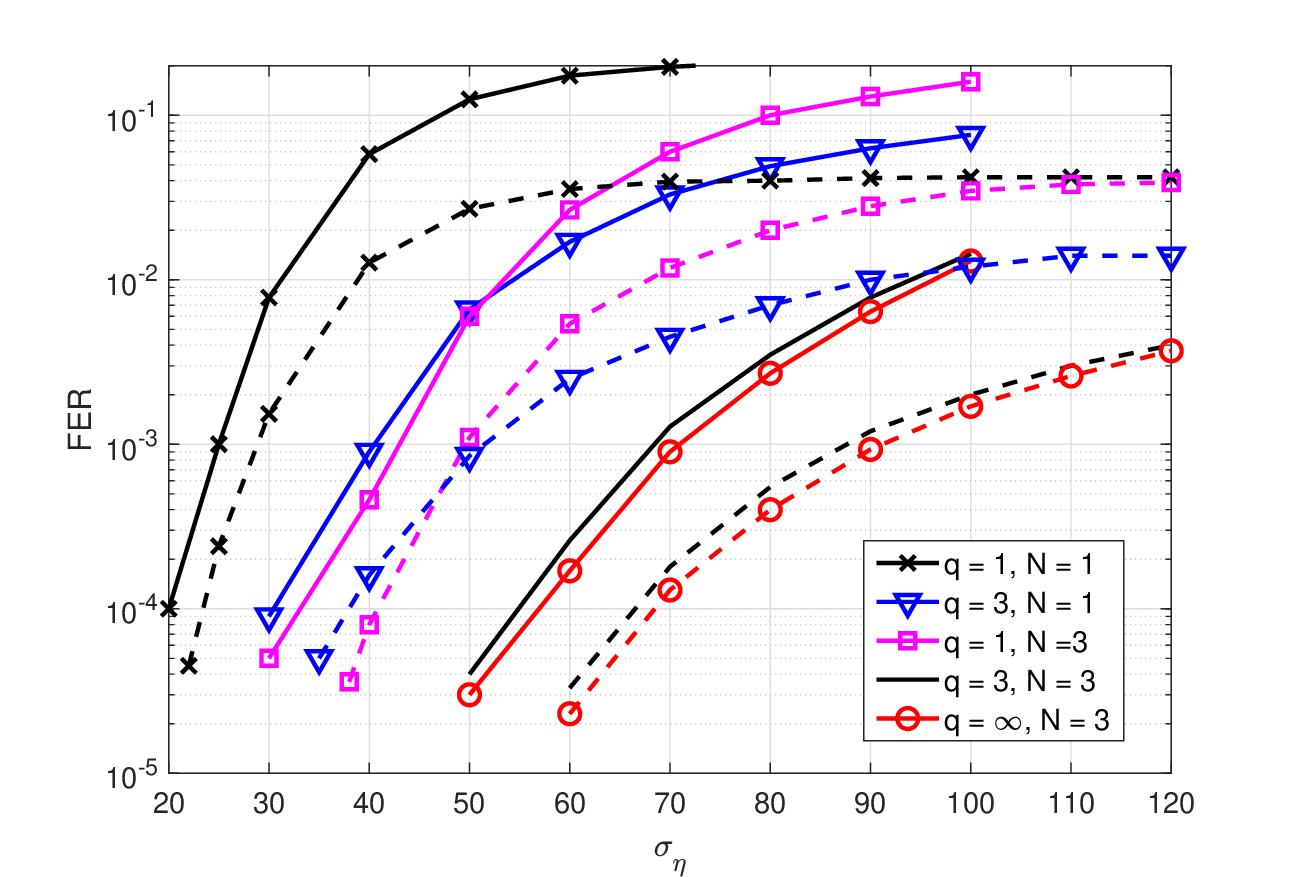}
\caption{The FER of (128, 110) polar code (solid line) and (128, 100) polar code (dash line) with our proposed quantizer with different numbers of reads and quantization bits.}
\label{Polar_SCL_CRC_FER}
\end{figure}

Both the uncoded BER and coded frame error rate (FER) performance are evaluated using computer simulations. In the simulations, we follow the literature \cite{ben2019detection} and take an array size of $m=n=16$, and $p_f=0.001$. For comparison, Detector 3 in \cite{ben2019detection} is also included in our simulations. To reduce the complexity of the MAP detector, Detector 3 in \cite{ben2019detection} is a threshold detector similar to that defined by \eqref{r_bar}. First, Fig. \ref{Uncoded_BER} shows that with $N=2$, there is an obvious gap between the uncoded BER of Detector 3 in \cite{ben2019detection} and that of the MAP detector. When $N$ increases to 6, this gap becomes even larger. On the other hand, our optimized detector can achieve the performance of the MAP detector for both $N=2$ and $N=6$.

As a class of ECCs widely applied to the data storage systems, two Bose–Chaudhuri–Hocquenghem (BCH) codes with different error correction capability are employed to examine the error rate performance of the single-bit quantization with HDD. As illustrated by Fig. \ref{BCH_FER}, with both the (127, 113) and (127, 92) BCH codes, the FER performance of our proposed detector can be improved significantly by increasing the number of reads $N$. On the contrary, as $N$ increases, the performance of Detector 3 in \cite{ben2019detection} is only improved slightly. This is due to the reason that the detection threshold of Detector 3 in \cite{ben2019detection} is only optimal for the single-read detection, but it has not been fully optimized for multiple reads.

%Polar codes have shown excellent SDD performance with short block lengths, and the decoding complexity is also low.
To evaluate the error rate performance of the system with multiple-bit quantization, ECC with SDD is required. We adopt a $(128, 110)$ polar code and a (128, 100) polar code with the cyclic redundancy check (CRC)-aided successive cancellation-list (SCL) decoding algorithm. The number of CRC bits is 4 and the list size is 8. 
%During the decoding process, the CRC detector is performed after the SCL decoding to select the correct codeword that passes the CRC. 
The simulation results illustrated by Fig. \ref{Polar_SCL_CRC_FER} indicate that for both two polar codes, by either increasing the number of quantization bits or the number of reads, the FERs can be greatly improved. Particularly, by using only 3-bit quantization and with $N=3$, we can approach the performance of polar codes with full SDD. Moreover, it is observed that single-bit quantization with 3 reads outperforms 3-bit quantization with single read at low FER regions ($\text{FER}<10^{-2}$). This indicates that, by applying more number of read operations, we can significantly improve the error rate performance with single-bit quantization and approach the performance with multiple-bit quantization.

%On the other hand, by taking into account the noise estimation error, we also simulate the FER of the polar code with $10\%$ underestimate of $\sigma_{\eta}$. As shown by Fig. 5,  the FERs with noise estimation error can still approach that without estimation error, which indicates that the our proposed quantization design method is very robust.

%With $N=8$ reads, the FERs the single-bit quantization outperforms that of the 3-bit quantization with $N=3$ reads. These observations are consistent with those obtained from Fig. \ref{MI_figure}.

\section{Conclusion}
We have proposed and optimized the MI-based single-bit and multiple-bit quantization with multiple reads for ReRAM. Both numerical and simulation results show that our proposed quantization schemes have approached the unquantized performance with only 3-bit quantization. For single-bit quantization, our proposed detector can achieve the performance of the MAP detector, which greatly outperforms the multiple-read detection in \cite{ben2019detection}. Moreover, simulation results reveal that increasing the number of reads can be more beneficial than increasing the number of quantization bits for the ReRAM channel, since it can provide more performance gain without the needs of SDD for ECCs.

\small
\bibliographystyle{IEEEtran}
\bibliography{postdoc_refs}

\begin{thebibliography}{10}
\providecommand{\url}[1]{#1}
\csname url@samestyle\endcsname
\providecommand{\newblock}{\relax}
\providecommand{\bibinfo}[2]{#2}
\providecommand{\BIBentrySTDinterwordspacing}{\spaceskip=0pt\relax}
\providecommand{\BIBentryALTinterwordstretchfactor}{4}
\providecommand{\BIBentryALTinterwordspacing}{\spaceskip=\fontdimen2\font plus
\BIBentryALTinterwordstretchfactor\fontdimen3\font minus
  \fontdimen4\font\relax}
\providecommand{\BIBforeignlanguage}[2]{{%
\expandafter\ifx\csname l@#1\endcsname\relax
\typeout{** WARNING: IEEEtran.bst: No hyphenation pattern has been}%
\typeout{** loaded for the language `#1'. Using the pattern for}%
\typeout{** the default language instead.}%
\else
\language=\csname l@#1\endcsname
\fi
#2}}
\providecommand{\BIBdecl}{\relax}
\BIBdecl

\bibitem{yu2016emerging}
S.~Yu and P.-Y. Chen, ``Emerging memory technologies: recent trends and
  prospects,'' \emph{IEEE Solid State Circuits Mag.}, vol.~8, no.~2, pp.
  43--56, Jun. 2016.

\bibitem{mei2018magn}
Z.~Mei, K.~Cai, and B.~Dai, ``Polar codes for spin-torque transfer magnetic
  random access memory,'' \emph{IEEE Trans. Magn.}, Nov. 2018.

\bibitem{strukov2008missing}
D.~B. Strukov, G.~S. Snider, D.~R. Stewart, and R.~S. Williams, ``The missing
  memristor found,'' \emph{nature}, vol. 453, no. 7191, p.~80, 2008.

\bibitem{cassuto2016information}
Y.~Cassuto, S.~Kvatinsky, and E.~Yaakobi, ``Information-theoretic sneak-path
  mitigation in memristor crossbar arrays,'' \emph{IEEE Trans. Inf. Theory},
  vol.~62, no.~9, pp. 4801--4813, Sep. 2016.

\bibitem{ben2019detection}
Y.~Ben-Hur and Y.~Cassuto, ``Detection and coding schemes for sneak-path
  interference in resistive memory arrays,'' \emph{IEEE Trans. Commun.}, Jun.
  2019.

\bibitem{chen2018coding}
Z.~Chen, C.~Schoeny, and L.~Dolecek, ``Coding assisted adaptive thresholding
  for sneak-path mitigation in resistive memories,'' in \emph{Proc. IEEE ITW},
  Nov. 2018.

\bibitem{song2021performance}
G.~Song, K.~Cai, X.~Zhong, Y.~Jiang, and J.~Cheng, ``Performance limit and
  coding schemes for resistive random-access memory channels,'' \emph{IEEE
  Trans. Commun.}, vol.~69, no.~4, pp. 2093--2106, Apr. 2021.

\bibitem{nguyen2021two}
C.~D. Nguyen, V.~K. Vu, and K.~Cai, ``Two-dimensional weight-constrained codes
  for crossbar resistive memory arrays,'' \emph{IEEE Commun. Lett.}, vol.~25,
  no.~5, pp. 1435--1438, May 2021.

\bibitem{chen2019variability}
Z.~Chen and L.~Dolecek, ``Variability-aware read and write channel models for
  1s1r crossbar resistive memory with high wordline/bitline resistance,''
  \emph{arXiv preprint arXiv:1912.02963}, 2019.

\bibitem{zorgui2019non}
M.~Zorgui, M.~E. Fouda, Z.~Wang, A.~M. Eltawil, and F.~Kurdahi,
  ``Non-stationary polar codes for resistive memories,'' in \emph{Proc. IEEE
  GLOBECOM}, Dec. 2019, pp. 1--6.

\bibitem{cai2015data}
Y.~Cai, Y.~Luo, E.~F. Haratsch, K.~Mai, and O.~Mutlu, ``Data retention in
  \text{MLC} \text{NAND} flash memory: Characterization, optimization, and
  recovery,'' in \emph{Proc. IEEE HPCA}, 2015.

\bibitem{tal2013construct}
I.~Tal and A.~Vardy, ``How to construct polar codes,'' \emph{IEEE Trans. Inf.
  Theory}, vol.~59, no.~10, pp. 6562--6582, Jul. 2013.

\bibitem{he2021dynamic}
X.~He, K.~Cai, W.~Song, and Z.~Mei, ``Dynamic programming for sequential
  deterministic quantization of discrete memoryless channels,'' \emph{IEEE
  Trans. Commun.}, Jun. 2021.

\bibitem{storn1997differential}
R.~Storn and K.~Price, ``Differential evolution--a simple and efficient
  heuristic for global optimization over continuous spaces,'' \emph{Journal of
  global optimization}, vol.~11, no.~4, pp. 341--359, 1997.

\end{thebibliography}

\end{document}